# Topological optical and phononic interface mode by simultaneous band inversion

O. Ortiz, P. Priya, A. Rodriguez, A. Lemaitre, M. Esmann[+] and N. D. Lanzillotti-Kimura[*]

*Université Paris-Saclay, CNRS, Centre de Nanosciences et de Nanotechnologies (C2N), 91120 Palaiseau, France*

*Interface modes have been widely explored in the field of electronics, optics, acoustics and nanophononics. One strategy to generate them is band inversion in one-dimensional superlattices. Most realizations of this type of topological states have so far been explored for a single kind of excitation. Despite its potential in the manipulation and engineering of interactions, platforms for the simultaneous topological confinement of multiple excitations remain an open challenge. GaAs/AlAs heterostructures exhibit enhanced optomechanical interactions due to the intrinsic colocalization of light and sound. In this work, we designed, fabricated, and experimentally studied a multilayered structure based on GaAs/AlAs. Due to the simultaneously inverted band structures for light and phonons, colocalized interface modes for both 1.34 eV photons and 18 GHz phonons appear. We experimentally validated the concept by optical reflectivity and coherent phonon generation and detection. Furthermore, we theoretically analyzed the performance of different topological designs presenting colocalized states in time-domain Brillouin scattering and deduce engineering rules. Potential future applications include the engineering of robust optomechanical resonators, compatible with the incorporation of active media such as quantum wells and quantum dots.*

## Introduction

As in the case of Fabry-Perot resonant modes, interface modes can be relevant for the development of Brillouin lasers and sensing applications. Interface modes have been widely explored in the field of electronics, optics, acoustics and nanophononics. In contrast to Fabry-Perot resonator modes, systematic engineering principles for interface modes were developed only much more recently. In this work, we address the problem of constructing a simultaneous optical and phononic interface mode by band inversion.

The concept of inverted spatial mode symmetries in periodic lattices often referred to as band inversion, is one of the cornerstones in the field of topological matter. Historically, this concept was first formalized with a comprehensive explanation for the electrical conduction properties of specific polymers using the Su-Schrieffer-Heeger (SSH) model[1–3]. Band inversion is used for the generation of an interface mode inside the bandgap by concatenating two periodic lattices with inverted bands. The existence of this interface state is protected against any perturbation that does not change the underlying topological invariants of the structures. In other words, the mode is pinned at the bandgap center due to the chirality of the system. The robustness of these topologically induced interface modes has been exploited in a wide range of physical systems[4–8] (photons[6,9–12], plasmons[4], phonons[7,8,13], vibrations[5,14], polaritons[15,16], classical acoustics[5,17–22]).

In this work, we report on a platform for the robust colocalization of light and sound on the nanoscale. To realize a system that presents band inversion for two excitations, we rely on GaAs/AlAs heterostructures that have been established as a platform for enhanced optomechanical interactions due to the intrinsic colocalization of the light and sound fields.

We designed, fabricated, and experimentally studied a heterostructure based on GaAs/AlAs presenting a topologically confined mode for both 1.34 eV photons and 18.12 GHz phonons. The topological mode is generated by concatenating two superlattices presenting inverted band structures.

## Concept and spectral characterization

An infinite, periodic superlattice is characterized by a folded dispersion relation where gaps are opened at the center and the edge of the Brillouin zone. For light the dispersion relation is governed by the indices of refraction. For acoustic phonons it is governed by the elastic constants of the materials.

A finite size superlattice can be associated to the infinite system through the correspondence between stop bands and band gaps, and the correspondence between reflection phases and Bloch mode symmetries. The finite system is usually defined as Distributed Bragg Reflector (DBR). For the sake of simplicity, we will use the terminology for the finite and infinite system interchangeably.

An interface mode can be systematically generated through band inversion. To this end, we concatenate two DBRs with the following conditions: 1. They share a common bandgap (same central frequency, same bandwidth) 2. The order of Bloch mode symmetries at the band edges enclosing this band gap is inverted. Under these conditions, the reflection phases of the two DBRs present opposite signs. This effect is directly rooted in the opposite topological properties of the individual bands and can be expressed in terms of the Zak phase[5,7,21]. At the bandgap center, the phases are equal in magnitude, resulting in a zero roundtrip phase thus giving rise to a resonance.

We control the order of Bloch mode symmetries through the internal structure of the DBR unit cells. In the left panel of Fig. 1(a), we show the dispersion relation of acoustic phonons in a GaAs/AlAs DBR. We have defined an inversion-symmetric trilayer unit cell. We have introduced the parameter $\delta$, which describes the relative acoustic thicknesses of the layers. The thickness of the unit cell layers is calculated as $(1+\delta)\lambda_{GaAs}/8$, $(1-\delta)\lambda_{AlAs}/4$, $(1+\delta)\lambda_{GaAs}/8$ where $\lambda$ is acoustic wavelength in the corresponding material at a frequency of 18.12 GHz. In this particular case, $\delta=-0.5$. At the second bandgap, the Bloch mode at the lower (upper) edge is symmetric (antisymmetric). We achieve a band inversion in the second minigap by choosing an opposite sign of $\delta$. For $\delta=+0.5$ (right panel of Fig. 1 (a)) the dispersion relation is the same with the exception of the mode symmetries. For other values of $\delta$, the width of the bandgap changes and for the particular case of $\delta=0$, the second bandgap is closed (not shown here).

For the case of light, the corresponding band structures are shown in Fig. 1(b). In this case, the second bandgap appears around 1.34eV. Up to a scaling factor, the overall band structures look the same as for the case of acoustics. Usually, for a given structure the optical and acoustic $\delta$ are by definition different. For acoustics, $\delta$ depends on the ratio of the speeds of sound in the two materials while for optics, it is the ratio of the speeds of light. In alloys formed from AlAs and GaAs, these ratios are equal.[23,24] Therefore, the band inversion can be simultaneously achieved for the optical and phononic bands in a single structure.

We concatenate two DBRs with an inverted second bandgap. This originates two colocalized interface modes at the center of the optical (1.34 eV) and acoustic (18.12 GHz) bandgap, respectively. In Fig. 1(c) we show the acoustic displacement $|u(z)|^2$ (black) and optical field $|E(z)|^2$ (red) of the topological interface modes with a perfect mode overlap. The blue and green colors are a guide to the eye to distinguish DBRs on the left and right, respectively. The optical quality factor for the confined mode is typically Q~2000.

We have fabricated a GaAlAs heterostructure consisting of alternating $Al_{0.95}Ga_{0.05}As$ and GaAs layers by molecular beam epitaxy technique on a (001) GaAs substrate. The structure is composed of two DBRs formed by 14 (16) periods of 65.0 nm / 230.7 nm (195.1 nm / 76.9 nm) GaAs/$Al_{0.95}Ga_{0.05}As$ layers for the left (right). The different number of pairs in the two DBRs accounts for the different refractive index of air and the substrate on both sides of the sample. This heterostructure simultaneously confines an optical interface mode at an energy of 1.34 eV and an acoustic phonon interface mode at a frequency of 18.12 GHz. The sample was grown with a spatial gradient such that it presents position-dependent resonance energies.

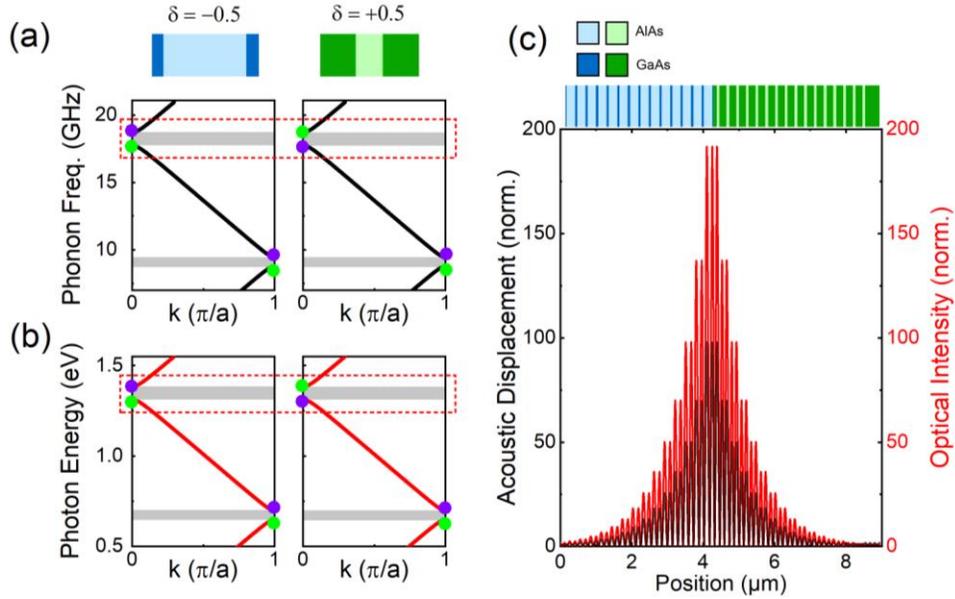

Fig. 1. Construction principle of an interface state in a photonic-phononic nanocavity by band inversion. (a) Schematic of the unit cells of each DBR parametrized by δ, which describes the relative thickness of the materials. Dark (light) colors represent GaAs (AlAs), blue and green are used to distinguish between left and right superlattices, respectively. The dispersion relations of acoustic phonons (a) and photons (b) present inverted Bloch mode symmetries around the first band gap at the Brillouin zone center. The dashed red box indicates the bandgap over which the band inversion takes place, mode symmetries are indicated with violet and green dots. (c) The spatial displacement pattern $|u(z)|^2$ (black) and the optical profile $|E(z)|^2$ (red) of the topological interface state at 18.12 GHz/1.34eV are colocalized and present a maximum at the interface between the two DBRs. Both fields decay evanescently away from the interface. The complete heterostructure is represented on top of panel (c).

To evidence the optical interface mode, we performed optical reflectivity measurements. Fig. 2(a) shows the reflectivity of the structure measured by FTIR (Bruker Equinox 55 with Hyperion Microscope, halogen lamp illumination). The two measured stop bands around 0.71 eV and 1.34 eV correspond to the first and second bandgap illustrated in Fig. 1(a). The dip in the middle of the second bandgap is associated with an optical cavity mode. Figure 2(b) provides a zoomed-in view of the interface state with a full width at half maximum (FWHM) of ~0.4 nm. The detail of this dip was measured using a home-built reflectometer based on a CW-laser (M2 SolsTiS). We have simulated the optical reflectivity using a transfer matrix formalism. The result (grey line) agrees with the experimental spectrum.

To acoustically characterize the structure and evidence the existence of a topological phononic interface mode, we performed time domain Brillouin scattering experiments[25,26] using an optical pump-probe technique. This approach can also evidence the dynamics between the confined acoustic and optical fields.

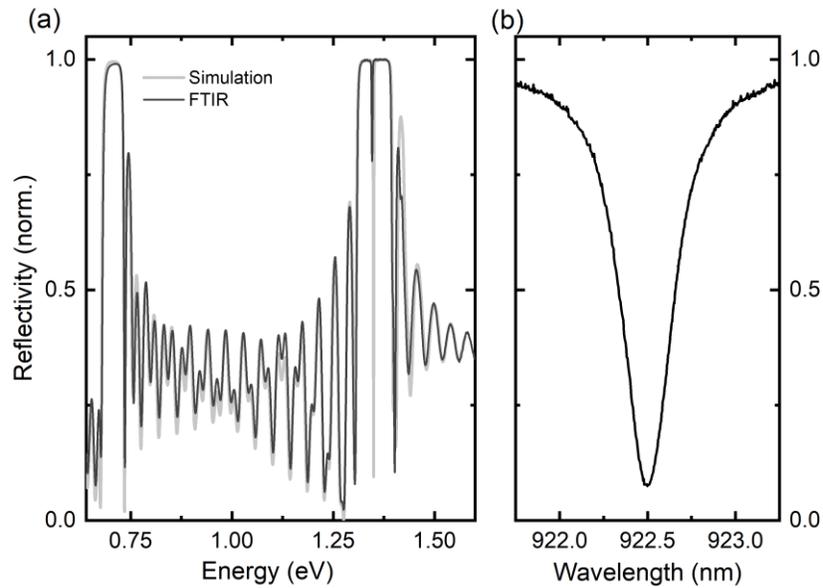

Fig. 2. Continous wave optical characterization of the topological interface modes. (a) The measured optical reflectivity reveals an interface state in the middle of the first optical stopband at the Brillouin zone center. (b) The zoomed-in optical reflectivity spectrum shows the interface state to be centered at 922.5 nm (1.34 eV).

We use a Titanium:Sapphire laser producing few picosecond long pulses at an 80 MHz repetition rate. The laser beam is split in two (pump and probe) following different paths, as shown in Fig. 3(a). The power of the pump and probe beams are 50 and 10 mW, respectively. The probe beam passes through a delay line to control the relative delay between pump and probe pulses. The pump beam goes through an acousto-optic modulator to allow a synchronous detection. Both beams are focused onto the sample into a spot of 5µm in a collinear geometry. The pump pulse impulsively generates coherent phonons with periods much longer than the pulse duration. These phonons modulate the optical properties of the sample with GHz frequencies. Using a cross-polarization scheme, a photodetector measures the resulting changes in the delay-dependent reflectivity experienced by the probe pulse[27,28]. A lock-in amplifier extracts differential reflectivity time traces like the ones shown in Fig. 3(b).

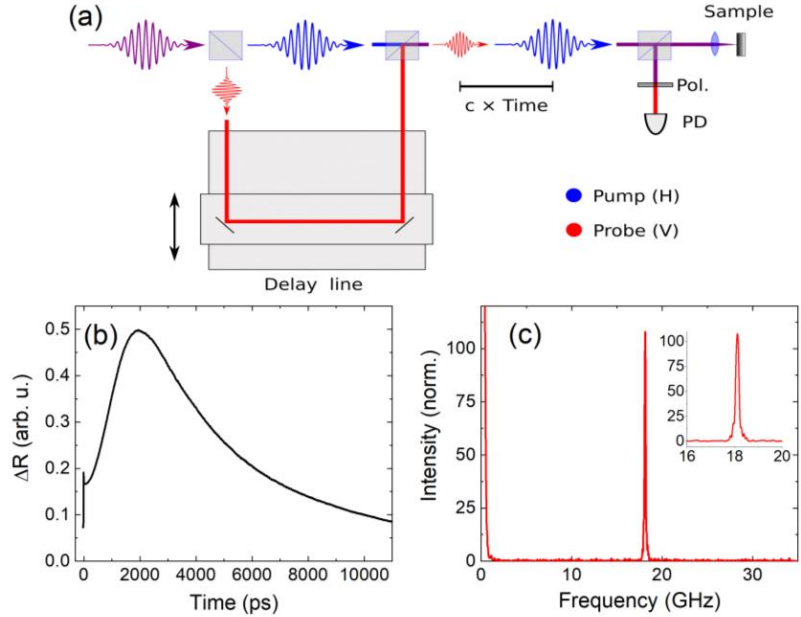

Fig. 3. Time-domain opto-phononic characterization of the topological interface mode. (a) Experimental coherent phonon generation setup. A mechanical delay line is used to control the delay between pump and probe pulses arriving on the sample. A cross-polarization scheme is used to probe the sample's instantaneous reflectivity while suppressing the reflected pump pulses. (b) Differential reflectivity time trace (0-11.000 ps) (c) Fast Fourier transform (FFT) of the experimental time trace featuring the topological resonator mode at 18.12 GHz. The inset shows a detail of the resonance with a full width at half maximum (FWHM) of 94 MHz.

In the time trace in Figure 3(b), the sharp change of reflectivity at t=0 is induced by the direct electronic excitation of the system. This change in reflectivity is followed by a slow variation where the signal reaches a maximum (~1 ns), and then it recovers its initial value (~11ns). These changes are induced by both the electronic dynamics and the temperature evolution of the sample.[28,29] A total of ~11 ns delay between pump and probe is scanned, covering approximately the full laser repetition period. The reflectivity trace within this window is Fourier transformed to obtain the acoustic spectrum of the topological resonator.

Figure 3(c) presents the Fourier transform of the derivative of the time trace shown in Figure. 3(b). We observe a clear peak at 18.12 GHz corresponding to the acoustic mode illustrated in Figure 1(b).

The inset to Figure 3(c) presents a zoomed-in spectrum of the resonance at 18.12 GHz, with a FWHM of 94 MHz. The FWHM can be mainly attributed to the finite length of the time trace (11ns corresponds to a Fourier limited linewidth of 90 MHz), the interference between phonons generated by two consecutive pump pulses, and lower signal detection efficiency for large delays.

In the time-resolved experiments, the pump pulse induces the excitation of the electronic system in the sample, generating a population of thermal phonons and coherent acoustic phonons. The thermal and electronic response of the structure induces a temporal evolution of the optical cavity, while the presence of the coherent acoustic phonons induces a periodic modulation of the index of refraction. The time-dependent resonance of the optical cavity determines a dynamical sensitivity to the presence of coherent phonons. The probe pulses measure the instantaneous optical reflectivity. The detection of phonons is also affected by the initial detuning of the pump and probe wavelength with respect to the optical cavity mode. The maximum generation of coherent acoustic phonons is reached when the pump laser is perfectly tuned with the cavity mode. The instantaneous reflectivity change is measured at the spectral position of the probe laser. The maximum detection sensitivity is reached when the probe laser is tuned to the highest reflectivity slope leading to the maximization of the signal.

Since the coherent phonons induce mainly a rigid shift of the cavity mode, the laser coupled to opposite sides of the optical mode leads to different signs of the phonon-induced change of reflectivity. In addition to the small variations induced by the coherent phonons (ps timescale)[25,26], the electronic excitation of the sample, and the thermal effects induce a slow change in the optical reflectivity mode (ns timescale).[29] The latter might induce an interesting dynamical effect: over the measuring time, the probe laser can pass through both slopes of the optical cavity mode, changing the sign of the phonon-induced changes.[28] We have further studied this phenomenon by tuning the laser power and wavelength across the cavity mode. The latter is easily implemented by taking advantage of the sample gradient.

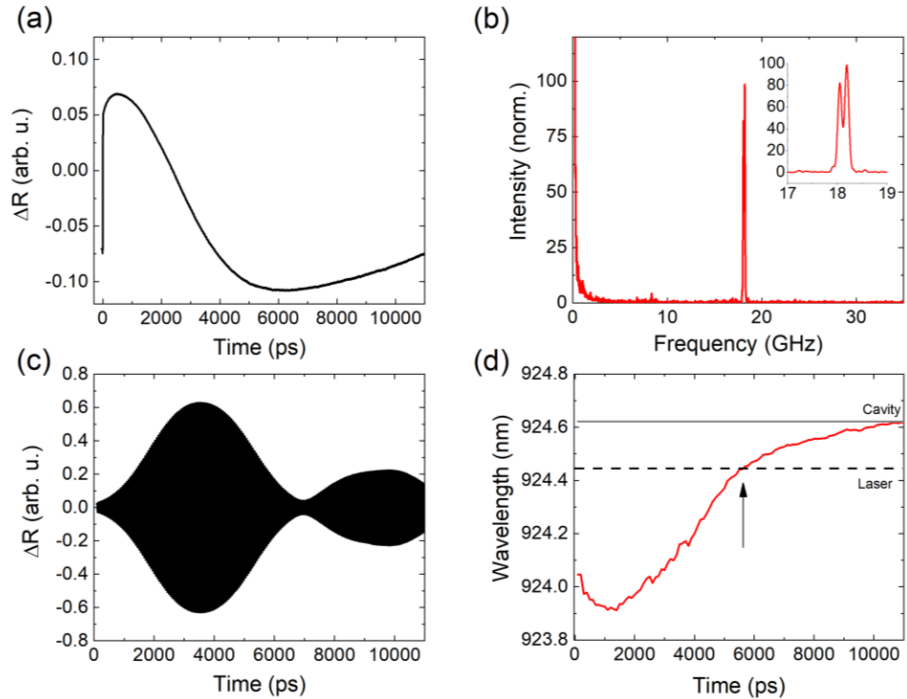

Fig. 4. (a) Differential reflectivity time trace (0-11.000 ps) taken at a different spectral detuning of the laser with respect to the cavity mode and different power setting. (b) Fast Fourier transform (FFT) of the experimental time trace featuring the topological resonator mode at 18.12 GHz. The inset shows a detail of the resonance which appears split with a separation of 140 MHz. (c) Time trace from panel (a) after passing a bandpass filter (17.97 GHz to 18.25 GHz). Coherent phonon oscillations are clearly visible with a modulated envelope. (d) The envelope in panel (c) is determined by the sensitivity of the structure to measure the modulation induced by the phonons. The red line indicates the position of the cavity as a function of delay, the dashed line the laser wavelength (fixed), the grey line indicates the unperturbed cavity resonance wavelength. The arrow shows the moment when the laser matches the cavity mode and the system is completely insensitive to phonons..

Figure 4(a) is the time trace obtained for a laser-cavity detuning of $\lambda_{Laser}-\lambda_{cav}$=-0.18 nm and a pump power of 50 mW. Note that for the data shown in Figure (3) these values are $\lambda_{Laser}-\lambda_{cav}$=-0.1 nm and 50 mW, respectively. In Fig. 4(a), there is a sharp change of reflectivity at t=0 followed by a slow variation where the signal reaches a maximum (~1 ns), then passes through a mimimum (~6ns) and and finally recovers its initial value (~12ns). Fig. 4(b) presents the Fourier Transform of the the time trace with the inset presenting a zoomed-in version of the resonance. Note that in contrast to Fig. 3(c) there are two peaks instead of one.

During the measurement, the probe passes by the minimum of reflectivity, causing the minimum in the trace near 6000 ps. Figure 4(c) presents the same time trace as in panel (a) after a bandpass filter between 17.97 GHz and 18.25 GHz. The envelope of the time trace is dominated by the sensitivity of the probe to the changes in reflectivity.

The evolution of the differential reflectivity can be directly associated to evolution of the optical cavity mode over time, as shown in Figure 4(d). We observe that the laser is tuned to both flanks of the cavity mode. The system returns to its equilibrium position for large delays.

Other parameters globally affecting this envelope are the acoustic-cavity lifetime and the divergence of the probe beam.[30] The carrier is the main contribution of the coherent acoustic phonons at 18.12 GHz, as indicated by the corresponding spectrum presented in Fig. 3(c). The node in the filtered time trace of Fig. 4(c) indicates the passage through the cavity mode minimum with a zero slope.

The splitting of the peak in Fig. 4(b) is a strong proof of phonon coherence since it constitutes a time-domain interference. This splitting can be modeled by considering the Fourier transform (FT) of a periodic signal (s) as a function of time delay (t) with an amplitude modulation such as $s(t) = \cos(\omega_{amp} t) \cdot \exp[-i \omega_{phonon} t]$. The resulting function $FT(s(t)) = S(\omega) = \sqrt{\pi/2} \cdot (\delta(\omega - (\omega_{phonon} - \omega_{amp})) + \delta(\omega - (\omega_{phonon} + \omega_{amp})))$ shows a splitting of $2\omega_{amp}$ around the resonant frequency $\omega_{phonon}$. This simple argument supports our image of the dynamic interaction between the photonic topological mode, determining the sensitivity to the present phonons, and the topological phononic mode, which is at the origin of the modulated signal in the GHz regime.

We have demonstrated the existence of the optical interface mode by optical reflectivity measurements. We showed the presence of the acoustic topologically confined mode by time-resolved Brillouin scattering and we discussed how they dynamically interact under specific excitation conditions. In the next section, we numerically investigate the efficiency in the Brillouin signal generation of different topological interface modes.

## Brillouin cross-section efficiency

We showed a possible configuration in which two concatenated superlattices with inverted bands give rise to an opto-phononic interface mode. By following the same principle, we can define a full class of resonators, which we explore in this section. The structures that we consider consist of two DBRs presenting inverted bands, and centrosymmetric unit cells. A band inversion can be implemented by two different operations: i) by changing the sign of delta or ii) by changing the superlattice inversion centers that coincide with the origin of the unit cells. That is, the materials of the layers forming a centrosymmetric unit cell are swapped. For all the discussion that follows, we consider resonators formed by two superlattices of 16 periods each one, surrounded by GaAs.

We present four particular topological structures corresponding to cases for the left and right DBRs with different values for **δ** and material composition of the symmetric unit cell. (a) is the same case, as presented in Fig. 1. (b) **δ**<0 for the left DBR and **δ**>0 for the right DBR, each one composed of unit cell centered on GaAs. (c) **δ**>0 and (d) **δ**<0 for both left and right DBRs constituted with symmetric unit cells centered on GaAs and AlAs, respectively. We also present the particular case of non-centrosymmetric unit cells formed by bilayers of GaAs/AlAs. This case (e) results in a conventional Fabry-Perot resonator formed with two identical DBRs composed of 3**λ**/4 (AlAs) and **λ**/4 (GaAs) layers enclosing a **λ**/2 spacer. Figure 5(a-e) displays the normalized acoustic displacement (black) and optical intensity (red) patterns of the interface modes.

In one-dimensional geometries, the frequency-dependent coherent phonon generation cross-section can be approximated by an overlap integral of the phononic strain ($\partial u/\partial z$), the electric field intensity $|E(z)|^2$ and the material-dependent photoelastic constant p(z).[31] We consider a photoelastic constant of $p_{GaAs}=1$ for GaAs layers and $p_{AlAs}=0$ for AlAs. For each structure, we have plotted the value of the integrand of the overlap integral for the photoelastic interaction (see Fig. 5(f-j)). The cross-section is determined by two conditions: i) the overlap between the antinodes of the opto-phononic field $|E(z)|^2(\partial u/\partial z)$ and the maxima of the photoelastic constant distribution p(z); ii) the relative sign of the optophononic antinodes in the regions where the photoelastic constant is non-zero. In addition, it is also important to consider that the antinodes with maximum amplitude at the interface between the two superlattices contribute the most to the overall cross-sections.

The integrands in Fig.5(f-j) can be analyzed by quadrants (left/right superlattice, positive/negative contributions). The signal in each quadrant is either formed by single or double peaks (thin/thick lines in the plot). The four cases confining phonons and light through band inversion Fig. 5(a-d) fall into two categories: In cases (c-d) the positive signal formed by double peaks in one superlattice mostly compensates the negative signals formed by single peaks in both superlattices. This results in an overall small Brillouin cross-section σ. Cases (a-b) feature positive signals composed of double peaks in one superlattice and single peaks in the other one while only one superlattice contributes negative single peaks. Effectively, the negative peaks compensate half of the positive double peaks leaving an overall signal with positive single peaked contributions from both superlattices. The main difference between cases (a) and (b) is the material of the layers forming the interface. The GaAs in case (a) contributes to the cross-section while the AlAs in case (b) does not.

The maximum cross-section is reached for the case of the Fabry-Perot resonator in Fig. 5(e) (σ=2939). Note that in this case, the integrand is definite-positive all over the structure, that is, two quadrants with positive single peaks. Note that the particular case of a Fabry-Perot resonator formed by two DBRs separated by a λ spacer (not shown here) would result in a vanishing cross-section, where the integrand would present positive peaks for one superlattice and negative peaks for the other.

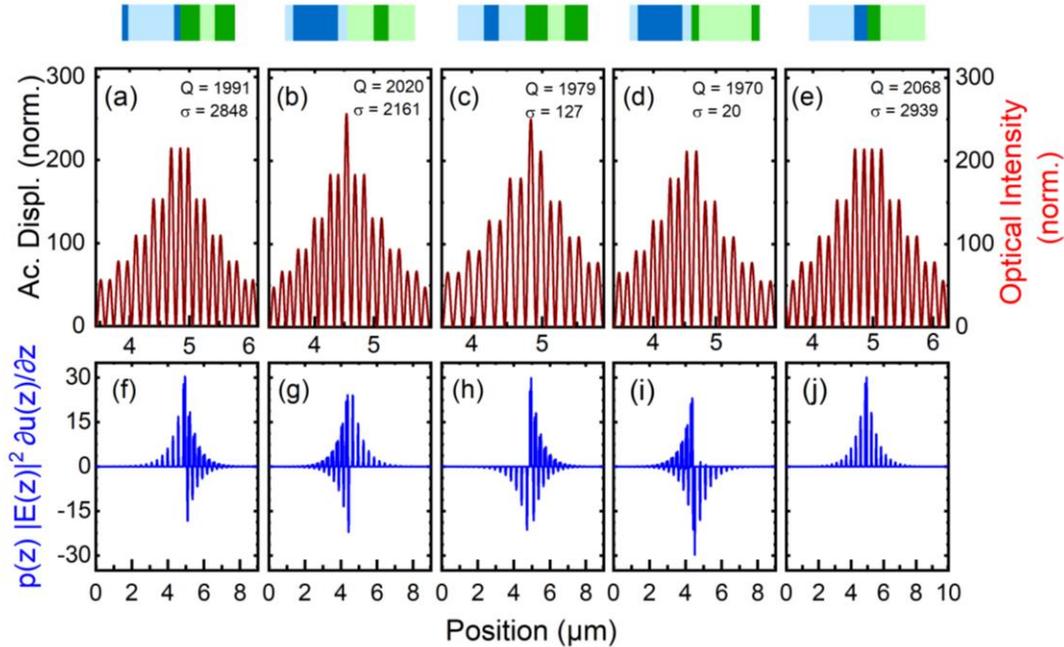

Fig. 5. Comparison between different topological structures and a Fabry-Perot resonator. The structures are designed to confine an acoustic mode at 18.12 GHz and an optical mode at 1.34 eV. (a-e) The spatial displacement pattern $|u(z)|^2$ (black) and optical profile $|E(z)|^2$ (red) are displayed for each structure. (a), (b), (c), (d) correspond to the topological structures composed of varying thicknesses and material composition, and (e) corresponds to a conventional Fabry-Perot resonator (see text for details). The top panel represents the unit cell of the DBRs at the interface of each structure. (f-j) Brillouin cross-section integrand (blue) corresponding to structures displayed in (a-e), respectively.

## Discussion and conclusions

Based on the band inversion concept applied to an optophononic GaAs/AlAs heterostructures, we conceived an interface mode where both NIR photons and GHz phonons are simultaneously confined. We reached Q-factors of the order of 2000 for both fields, which are phonons at 18.12 GHz and photons at 1.34 eV. The full structure thickness was of the order of 10 µm, reachable with standard MBE techniques. The GaAs/AlAs platform dictates the perfect overlap between the electric and the atomic displacement fields.

We performed coherent phonon generation experiments based on pump-probe differential reflectivity measurements. We observed acoustic-phonon lifetimes longer than 10 ns. The experimental signatures show the convoluted response of the optical and the acoustic cavity, unveiling the interplay between the two confined fields.

Usually, coherent control experiments[32–34] are necessary to prove the coherent character of the impulsively generated acoustic phonons. Here, the split peak in Fig. 4b is a direct proof of phonon coherence since it is due to interference between optical signals at different times, probing the same coherent acoustic phonons. The use of an optical cavity mode allows us to flip the phase of the detection in the middle of the measurement by just choosing the appropriate initial excitation conditions, pump power, and detuning.

Our optical experiments evidenced the existence of the optical and phononic topological interface modes. The chosen structure is not the only topological resonator that can be constructed in GaAs/AlAs heterostructures. By comparing different combinations of concatenated superlattices, we numerically showed that despite similar Q-factors, the Brillouin cross-section of the experimentally chosen structure is the highest, almost matching the performance of a topologically trivial Fabry-Perot resonator.

A significant advantage of the studied topological opto-phononic resonator is the stability of the modes with respect to chirality-preserving fluctuations in the layer thicknesses. The reported resonator not only presents robust optical and acoustic modes[5,7] but also a robust signal when comparing the performance with a standard Fabry-Perot resonator of similar characteristics. Eventhough the colocalization is preserved in both cases, the interaction presents a different dependence on these fluctuations. In the particular case of a Brillouin interaction we can consider two Gedankenexperiments: 1) The laser wavelength is fixed. In this case the stronger fluctuations of the cavity resonance for the Fabry-Perot resonator manifest themselves in pronounced intensity fluctuations of the Brillouin signal. This is not the case for any of the topological cavities. 2) The laser wavelength follows the position of the cavity mode. In this case the measured Brillouin spectrum from the Fabry-Perot resonator would be subject to stronger wavelength instabilities than the Brillouin spectrum of the topological optophononic resonator. Intensity fluctuations, however, would be similar for both types of resonators in this latter case.

A 3D control of the optical and acoustic densities of states could be achieved by etching micropillars out of the studied heterostructures. Potential future applications of the reported results thus include the engineering of optomechanical resonators, where the overlap of the optical and acoustic fields is a fundamental requirement. The use of standard III-V materials makes the system a potential testbed of novel concepts involving active media such as quantum wells and quantum dots, where the controlled of the light-matter interactions might prove to be a key asset.

## Funding

The authors acknowledge funding by the European Research Council Starting Grant No. 715939, Nanophennec. This work was supported by the European Commission in the form of the H2020 FET Proactive project TOCHA (No. 824140). The authors acknowledge funding by the French RENATECH network and through a public grant overseen by the ANR as part of the "Investissements d'Avenir" program (Labex NanoSaclay Grant No. ANR-10-LABX-0035). M.E. acknowledges funding by the Deutsche Forschungsgemeinschaft (DFG, German Research Foundation) Project 401390650.

## Author Contributions

NDLK and ME proposed the concept and designed the device. AL fabricated the sample. All the authors performed the experiments and simulations, discussed and analyzed the results and wrote the paper. NDLK guided the research.